\documentstyle[aps]{revtex}
\begin{document}
\draft
\preprint{SINP-98/05, hep-th 9802111}
\title{Gaugino mass in the Heterotic string with Scherk-Schwarz compactification}
\author{Parthasarathi Majumdar$^{\dagger}$\footnote{email: partha@tnp.saha.ernet.in,
partha@imsc.ernet.in}  and Soumitra
SenGupta$^{\ddagger}$\footnote{soumitra@juphys.ernet.in}}
\address{$^{\dagger}$ Saha Institute of Nuclear Physics, AF/1
Bidhannagar, Calcutta 700 064, India and \\ The Institute of Mathematical Sciences,
Chennai 600 113, India (permanent address). \\ $^{\ddagger}$ Physics Department,
Jadavpur University, Calcutta 700 032, India.}
\maketitle
\begin{abstract}
The generic observable sector gaugino mass in the weakly-coupled heterotic string
compactified to four dimensions by the Scherk-Schwarz scheme (together with hidden
sector gaugino condensation 
inducing the super-Higgs effect with a vanishing cosmological constant) is shown to
be
nonzero at tree level, being of the order
of the gravitino mass, modulo reasonable assumptions regarding the magnitude of the
condensate and the Scherk-Schwarz mass parameters. 
\end{abstract}

Despite possessing ingredients of phenomenological relevance, the weakly coupled $E_6
\times E_8$ heterotic string theory suffers from the one key lacuna, viz., a
non-perturbative mechanism of breaking spacetime supersymmetry with a vanishingly small
cosmological
constant. Conventionally, low energy approximations in four dimensions
(on toroidal Kaluza-Klein compactification\cite{wit}) are derived
assuming a condensation of the gauginos of the `hidden sector' $E_8$ super-Yang Mills
theory \cite{nil}, and compensating the resulting vacuum energy by ascribing a
finely-tuned expectation
value to the internal components of the antisymmetric tensor field \cite{rom}. The
resulting supergravity theory undergoes super-Higgs effect with generation of
a gravitino mass \cite{iban}, \cite{brei}. This, however, precludes a tree
level mass for the generic observable gaugino \cite{brig}. 

To see this, recall that the gaugino mass is given, in $D=4, N=1$ supergravity, by
the general formula \cite{crem} (in Planckian units)
\begin{equation}
(m_{\frac12})_{ab}~=~\langle \left [ e^{{\cal G}/2}~ {\cal G}^l~({\cal G}^{-1})_l~^k
\right ]
f_{ab,k} \rangle ~\label{mggn}
\end{equation}
where, for the $E_6 \times E_8$ heterotic string, the supergravity functions ${\cal
G}$ and $f_{ab}$ of the moduli and matter scalar fields are given by \cite{nil}
\begin{equation}
{\cal G}~=~\log \left(~|W|^2 \over {(S+{\bar S}) (T+{\bar T} -2 |C_i|^2)^3
} \right)~,~f_{ab}~=~S~\delta_{ab}~,~\label{torgf} \end{equation}
with $S~,~T$ being $E_6$ singlet moduli fields related to the dilaton and the
axion, while the $C_i ({\bar C}^i)$ are 27-plet (${\bar {27}}$)-plet matter fields, and
$W$ is the superpotential, given by
\begin{equation}
W~~=~~\lambda^{ijk} C_i C_j C_k~+~W(S)~. \label{spot} \end{equation}
Here, $W(S)$ is the {\it effective} superpotential subsuming the effect of 
gaugino condensation in the hidden sector super-Yang Mills theory, as
discussed by \cite{nil}, \cite{brei}. Its structure will be exhibited in the sequel. 
Minimizing the resulting scalar potential and
requiring that it vanishes at its minimum yields the vacuum conditions \cite{brei}
\begin{equation}
\langle C_i \rangle~=~0~=~\langle {\cal G}_S \rangle~. \label{vac} \end{equation}
The gravitino mass arising as a consequence is given by
$m_{\frac32}~=~\langle e^{{\cal G}/2} \rangle ~ \neq 0$, so that, the ratio
$m_{\frac12} / m_{\frac32}$ vanishes as a result of the vacuum conditions (\ref{vac}).
Radiative corrections scarcely remedy the malady vis-a-vis phenomenological
compulsions. 

While the fundamental enigma continues, a manner in which the
theory may be made to look more phenomenologically presentable was pointed
out several years ago \cite{por}. This approach replaces standard toroidal
compactification of the weak coupling heterotic string by the alternative
route pioneered by Scherk and Schwarz \cite{sch} in the context of
supergravity. The set of chiral superfields appearing in the effective
$D=4, N=1$ supergravity now includes, over and above $S$, the matter
27-plets $\Phi^{i_A}~,~A=1,2,3$, ${\bar {27}}$-plets ${\bar \Phi}_{i_A}$,
singlet moduli $U^A, {\bar U}_A$ and $T_A, T'_A$, all of which are
collectively designated as ${\cal Y}^{I_A}$. In terms of these fields, the
effective $D=4,N=1$ supergravity is given by the functions
\begin{equation}
{\cal G}~=~\log \left (~{ |{\tilde W}|^2 \over {Y_0 Y_1 Y_2 Y_3}}
~\right)~,~ f~=~S~. ~\label{ssgf} \end{equation}
Here, $Y_0 ~\equiv~2 ReS $,
\begin{equation}
Y_A~\equiv~1~-~|{\cal Y}^{I_A}|^2~+~\frac12 |({\cal Y}^{I_A})^2|^2
~,\label{wyi} \end{equation}
and 
\begin{equation}
{\tilde W}~\equiv~d_{i_A j_B k_C} \Phi^{i_A} \Phi^{j_B}
\Phi^{k_C}~+~\Delta^{SS} W~, \label{sssp} \end{equation}
with 
\begin{eqnarray}
\Delta^{SS} W & \equiv & \frac{m_1}{2} \left (1 + \sqrt{2} T_1 + \frac12
({\cal
Y}^{I_1})^2 \right) \left [ \left( 1-\sqrt{2} T_2 +
\frac12 ({\cal Y}^{I_2})^2 \right ) \left ( 1+ \sqrt{2} T_3 + \frac12
({\cal Y}^{I_3})^2 \right ) + \left( 2 \leftrightarrow 3 \right ) \right]
\nonumber \\
&-& \frac{m_2}{2} \left( 1+ \sqrt{2}T_1 + \frac12 ({\cal Y}^{I_1})^2
\right) \left [ \left( i - \sqrt{2} T'_2 - \frac{i}{2} ({\cal
Y}^{I_2})^2 \right) \left( i + \sqrt{2} T'_3 - \frac{i}{2} ({\cal Y}^{I_3})^2
\right) + \left( 2 \leftrightarrow 3 \right) \right] ~.   \label{sspot} 
\end{eqnarray}  

The scalar potential has several flat directions in this case; the
simplest choice for the vacuum is given by \cite{por} $< {\cal Y}^{I_A}
>~=~0$. Now if we require the vacuum energy to vanish for finite
non-vanishing $<Y_0>$, it turns out in this case that \cite{ssg} one must
choose the Scherk-Schwarz mass parameters $m_1~,~m_2$ to obey
$m_1~+~m_2~=~0$. It
follows that the gravitino mass 
\begin{equation}
m_{\frac32}~\equiv~<e^{{\cal G} /2}>~=~{ {m_1~+~m_2} \over <
Y_0>^{\frac12}}~=~0~, \label{gnoss}
\end{equation}
implying that, although the Scherk-Schwarz mechanism does break
supersymmetry in principle, in the present situation, the requirement
of a vanishing
cosmological constant is too stringent for this breaking to survive
\cite{ssg}. Using eq.s (\ref{mggn}), (\ref{ssgf}) - (\ref{gnoss}), it is
easy to see that in this case, $m_{\frac12}~=~0$, as of course is expected on
grounds of consistency. To effect a genuine supersymmetry breaking vacuum
structure, one needs a non-perturbative phenomenon, like hidden
sector gaugino condensation as discussed earlier, to occur concomitantly  
with the Scherk-Schwarz mechanism \cite{ssg}. Before turning to such a
hybrid scenario,
which we do next, we observe {\it en passant} that, were we to ignore
the restrictions of a vanishing cosmological constant, we would get, for
the vacuum choice made above, $m_{\frac12} \sim m_{\frac32}$. 

A fusion of the Scherk-Schwarz mechanism with hidden sector gaugino condensation
in the weak coupling heterotic string has already been considered in \cite{ssg},
\cite{sspm}. Here
we focus on the implications of such a marriage on the generic observable gaugino.
We follow ideas of ref.s \cite{nil} and \cite{brei} to stipulate that the primary
outcome of
gaugino condensation in the hidden sector, is to augment the superpotential 
(\ref{sssp}) as follows
\begin{equation}
{\tilde W} ~~\rightarrow~~{\cal W}~~\equiv~~{\tilde W}~+~W(S)~.\label{ssp'}
\end{equation}
Here, $W(S)$ is the effective superpotential for the modulus $S$, possessing the
generic structure \cite{brei}
\begin{equation}
W(S)~~\sim~~{\cal A}~+~{\cal B}~ \exp -3 (~2 \sigma~+~S/2b_0~)~,\label{ws}
\end{equation}
with ${\cal A}, {\cal B}$ constants, $\sigma$, the `breathing mode' arising in
the compactification and $b_0$, the coefficient of the one loop beta function of the
hidden sector gauge group (for $E_8$, $b_0=60$).  

The corresponding scalar potential once again has a minimum with vanishing 
cosmological constant \cite{ssg}; this is given by the vacuum conditions
\begin{equation}
\langle{\cal G}_S \rangle ~=~0~=~\langle {\cal Y}^{I_A}
\rangle~\rightarrow~\langle {\cal G}_{i_A}\rangle~=~\langle{\cal
G}_A\rangle~=~0~.\label{vacss}
\end{equation}
However, 
\begin{eqnarray}
\langle~{\cal G}^A~\rangle~&=&~\langle~{{\cal W}^A \over {\cal W}}~\rangle~=~{{(m_1 + m_2)~\delta_1^A} \over
{(m_1 + m_2 +
\langle W(S) \rangle )}}~\nonumber \\
 \langle{\cal G}^{A'}\rangle~&=&~\langle{{\cal W}^{A'} \over {\cal W}}\rangle~=~{{(m_1
+ m_2)~\delta_1^{A'}}
\over {(m_1 + m_2 +\langle W(S) \rangle)}}~.\label{wa} \end{eqnarray}
Since in this case, the requirement of a vanishing cosmological constant does not
constrain the Scherk Schwarz mass parameters $m_1, m_2$, thereby allowing a
non-vanishing gravitino mass $m_{\frac32}~=~\langle e^{{\cal G}/2} \rangle$, at least
for
the particular point in moduli space given by (\ref{vacss}). Using once again eq.s
(\ref{mggn}), (\ref{ssgf})-(\ref{gnoss}) and (\ref{ssp'})-(\ref{wa}), it is
straightforward to obtain the ratio
\begin{eqnarray}
{m_{\frac12} \over m_{\frac32}}~~&=&~~(m_1~+~m_2)^2~\langle {W_S(S) \over {\cal W}^3}
\rangle
\nonumber \\
&=&~~{{(m_1~+~m_2)^2~\langle W_S(S) \rangle} \over { [m_1~+~m_2~+~\langle W(S)
\rangle]^3 \langle {\cal G}^2 \rangle}}~.\label{ggno} \end{eqnarray}
This establishes that the generic gaugino mass no longer vanishes at tree level, in
contrast to what ensued in the case of the toroidally compactified heterotic string. 

An order of magnitude estimate of the rhs of eq. (\ref{ggno}) can be obtained upon
using the formula \cite{brei}
\begin{equation}
\langle Y_0 \rangle^{-1}~~=~~{g^2 \over {4\pi}}~, \label{yo} \end{equation}
where $g$ is the value of the gauge coupling constant of the hidden sector (super-)
Yang Mills theory around Planck scale. The constant ${\cal B}$ in (\ref{ws}) can be
estimated by identifying the vacuum expectation value $\langle W_S(S) \rangle$ with
the hidden sector gaugino condensate, yielding ${\cal B} \sim -\frac{1}{3} b_0$. We now
make the `reasonable' choices 
\begin{equation}
{\cal A}~ \sim ~-{\cal B}~ \exp  -3 [2 \langle \sigma \rangle + 1/b_0
g^2]~,~m_1~+~m_2~\sim~\exp-3 [2 \langle \sigma \rangle + 1/b_0
g^2]~, \label{chcs} \end{equation}
recalling that all dimensional quantities are in Planckian units. The parameters
$\langle \sigma \rangle$ and $g$ may now be fine tuned to produce a ratio
$|m_{\frac12}/m_{\frac32}| ~\sim ~1$. 

A few disclaimers are in order. First of all, the foregoing does not constitute
anything beyond a `feasibility study' of a non-vanishing tree level gaugino mass of a
desirable magnitude within the compound scenario incorporating both Scherk Schwarz
compactification and hidden sector gaugino condensation. Since our assumptions involve
quantities arising in a strongly interacting gauge theory whose low energy dynamics is
by no means well-understood, numerical estimates of quantities appearing above are not
meant to be taken too seriously. Secondly, as pointed out in ref. \cite{por}, the
vacuum chosen is but a point in the moduli space of the theory, and it is not claimed
that the estimate of the ratio above being of the order unity holds over the entire
space. Presently available technology, however, does not permit one to ascertain any
preferred point in moduli space, thus reinforcing the vacuum ambiguities well-known in
the case of toroidal compactification \cite{nil}, \cite{brei}. Thirdly, a complete
estimate of all possible soft operator coefficients (the scalar mass squared and the
trilinear scalar coupling constant) has to emerge consistently -- a problem to be soon
resolved, hopefully. Last, but definitely not least, the effect of stringy restrictions
on the Scherk Schwarz mechanism, as for instance those associated with the two
dimensional conformal invariance of the heterotic string discussed in
\cite{ps}  and \cite{sspm} have to be carefully (re)examined. It may well
turn out that these are so
stringent as to {\it inhibit} supersymmetry breaking with a vanishing cosmological
constant \cite{sspm}. 

Finally, the problem of a generic observable gaugino has recently received attention 
\cite{nilles}, \cite{ovru} within the {\it strongly} coupled $E_8 \times E_8$ heterotic
string (related to M-theory a l/'a Horava and Witten \cite{hor},)  using hidden sector
gaugino condensation. The
general problem of supersymmetry breaking in M-theory using Scherk Schwarz 
compactification is also under purview \cite{dud}. It may be of some interest to
consider this problem from the augmented standpoint adopted in this paper. We hope to
report on this elsewhere.

SS acknowledges illuminating discussions with P. Binetruy,
K. Dienes, E. Dudas, P. Fayet and N. Sakai at the SUSYUNI meeting in Mumbai,
India in December, 1997. PM thanks the Theory Group of Saha Institute of Nuclear
Physics for hospitality during a visit during which this work was completed.

 \end{document}